\documentclass[a4paper,11pt]{article}
\pdfoutput=1 

\usepackage{jheppub} 

\usepackage[T1]{fontenc} 
\usepackage{graphicx}

\usepackage{MnSymbol}

\usepackage{slashed} 

\title{\boldmath Tree-level  gluon amplitudes on the celestial sphere}

\author{Anders {\O}. Schreiber,$^1$}
\author{Anastasia Volovich,$^{1,2}$}
\author{Michael Zlotnikov$^1$}

\affiliation{$^1$ Department of Physics,
  Brown University,
  Providence RI 02912}

\affiliation{$^2$ School of Natural Sciences,
  Institute for Advanced Study,
  Princeton NJ 08540}

\emailAdd{anders\_schreiber@brown.edu}
\emailAdd{anastasia\_volovich@brown.edu}
\emailAdd{michael\_zlotnikov@brown.edu}

\abstract{\\
Pasterski, Shao and Strominger have recently proposed that massless scattering amplitudes can be mapped to correlators on the celestial sphere at infinity via a Mellin transform. We apply this prescription to arbitrary $n$-point tree-level gluon amplitudes. The Mellin transforms of  MHV amplitudes are given by generalized hypergeometric functions on the Grassmannian $Gr(4,n)$, while generic non-MHV amplitudes are given by  more complicated Gelfand $A$-hypergeometric functions.  

}

\begin{document} 
\maketitle
\flushbottom

\section{Introduction}
The holographic description of bulk physics in terms of a theory
living on the boundary has been concretely realised
by the AdS/CFT correspondence for spacetimes with global negative curvature.
It remains an important outstanding problem to understand suitable
formulations of holography for flat spacetime, a goal that
has elicited a considerable amount of work from several complementary
approaches \cite{flat}.

Recently, Pasterski, Shao and Strominger  \cite{Pasterski:2016qvg} studied the scattering of  particles in four-dimensional Minkowski space and formulated a prescription that maps these amplitudes to the celestial sphere at infinity. The Lorentz symmetry of four-dimensional Minkowski space acts as the conformal group $SL(2,\mathbb{C})$ on the celestial sphere. It has been shown explicitly that the near-extremal three-point amplitude in massive cubic scalar field theory has the correct structure to be identified as a three-point correlation function of a conformal field theory living on the celestial sphere \cite{Pasterski:2016qvg}. The factorization singularities of more general scattering amplitudes in this CFT perspective have been further studied in 
\cite{Cardona:2017keg,Nandan:2018}. The map uses conformal primary wave functions which have been constructed for various fields in arbitrary dimensions in \cite{Pasterski:2017kqt}. 
In  \cite{Pasterski:2017ylz} it was shown that the change of basis from plane waves to the conformal primary wave functions is implemented by a Mellin transform,
which was computed explicitly for three and four-point tree-level gluon amplitudes. The optical theorem in the conformal basis and
scattering in three dimensions were  studied in \cite{Lam:2017ofc}. One-loop and two-loop four-point amplitudes have also been considered in \cite{Banerjee:2017jeg}.

In this note  we use  the prescription \cite{Pasterski:2017ylz} 
to investigate the structure of CFT  correlators corresponding to arbitrary $n$-point gluon tree-level  scattering amplitudes, thus generalizing their three- and four-point MHV
results. Gluon amplitudes can be represented in many different ways that exhibit different, complementary aspects of their rich mathematical structure. It is natural to suspect that they may also take a particularly interesting form when written as correlators on the celestial sphere. We find that Mellin transforms of
$n$-point  MHV gluon amplitudes  are given by  Aomoto-Gelfand generalized hypergeometric functions  on the Grassmannian $Gr(4,n)$ (\ref{gluonPhi2}).
For non-MHV amplitudes the analytic structure of the resulting functions is more complicated, and they are given by Gelfand $A$-hypergeometric functions (\ref{NMHVFres}) and its generalizations. 
It will be very interesting to explore further the structure of these functions, and possibly make connections to other representations of tree-level amplitudes \cite{other}
which we leave for future work.

\section{Gluon amplitudes on the celestial sphere}
\label{gluonSec}
We work with tree-level $n$-point scattering amplitudes of massless particles $\mathcal{A}_{\ell_1\cdots \ell_n}(k_j^{\mu})$  which are functions of external momenta $k_j^{\mu}$ and  helicities $\ell_j=\pm 1,$ where $j= 1, \ldots, n$. We want to map these scattering amplitudes to the celestial sphere. To that end we can parametrize the massless external momenta $k_j^{\mu}$ as
\begin{align}\label{qvec}
k_j^\mu=\epsilon_j\omega_j q_j^\mu \equiv \epsilon_j\omega_j ( 1+|z_j|^2 , z_j+\bar z_j , -i (z_j-\bar z_j) , 1-|z_j|^2),
\end{align}
where $z_j, \bar{z_j}$ are the usual complex cordinates on the celestial sphere, $\epsilon_j$ encodes a particle as incoming ($\epsilon_j =-1$) or outgoing ($\epsilon_j = +1$), and $\omega_j$ is the angular frequency associated with the energy of the particle \cite{Pasterski:2017ylz}. Therefore, the amplitude 
$\mathcal{A}_{\ell_1\cdots \ell_n} (\omega_j,z_j,\bar z_j)$
is
 a function of $\omega_j$, $z_j$ and $\bar{z}_j$ under the parametrization (\ref{qvec}). 

Usually, we write any massless scattering amplitude in terms of spinor-helicity angle- and square-brackets representing Weyl-spinors (see \cite{Srednicki:2007qs} for a review). The spinor-helicity variables are related to external momenta $k_j^\mu$, so that in turn we can express them in terms of variables on the celestial sphere via \cite{Pasterski:2017ylz}:
\begin{align}
\label{spihel}
[ij] =  2\sqrt{\omega_i \omega_j} \bar z_{ij} \,,~~~~~~\langle ij\rangle = -2\epsilon_i\epsilon_j\sqrt{\omega_i \omega_j } z_{ij}\,,
\end{align}
where $z_{ij} = z_i - z_j$ and $\bar z_{ij} = \bar z_i - \bar z_j.$

In \cite{Pasterski:2017kqt, Pasterski:2017ylz} it was proposed that any massless  scattering amplitude   is mapped to the celestial sphere via a Mellin transform:
\begin{align}\label{mappingGen}
\mathcal{\tilde A}_{J_1\cdots J_n}(\lambda_j, z_j,\bar z_j)
=  \prod_{j=1}^n  \int_0^\infty d\omega_j \, \omega_j^{i\lambda_j } \,
\mathcal{A}_{\ell_1\cdots \ell_n} (\omega_j,z_j,\bar z_j)\,.
\end{align}
The Mellin transform maps a plane wave solution for a helicity $\ell_j$ field in momentum space to a corresponding conformal primary wave function on the boundary with spin $J_j$, where helicity $\ell_j$ and spin $J_j$ are mapped onto each other, and the operator dimension takes values in the principal continuous series representation $\Delta_j = 1 + i \lambda_j$ \cite{Pasterski:2017kqt}. Therefore, $\mathcal{\tilde A}_{J_1\cdots J_n}(\lambda_j, z_j,\bar z_j)$ has the structure of a conformal correlator on the celestial sphere, where the symmetry group of diffeomorphisms is the conformal group $SL(2, \mathbb{C})$.  

Explicitly, under conformal transformations, we have the following behavior:
\begin{align}
\label{confscale}
\omega_j\to\omega_j'=|cz_j+d|^2\omega_j~~~,~~~z_j\to z_j'=\frac{a z_j+b}{c z_j+d}~~~,~~~\bar z_j\to \bar z_j'=\frac{\bar a \bar z_j+\bar b}{\bar c \bar z_j+\bar d},
\end{align}
where $a,b,c,d\in\mathbb{C}$ and $a d-b c=1$. The transformation for $z_j,\bar z_j$ is familiar from the usual action of $SL(2, \mathbb{C})$ on the complex coordinates on a sphere. Concerning $\omega_j$, recall that $q_j^\mu$ transforms as $q_j^\mu \rightarrow |c z_j + d|^{-2} \Lambda^\mu{}_\nu q_j^\nu$ \cite{Pasterski:2017kqt}, where $\Lambda^\mu{}_\nu$ is a Lorentz transformation in Minkowski space corresponding to the celestial sphere conformal transformation. Thus, $\omega_j$ must transform as in  (\ref{confscale}) to ensure that $k^\mu_j$ transforms as a Lorentz vector: $k_j^\mu\to \Lambda^\mu{}_\nu k_j^\nu$. 

The conformal covariance of $\mathcal{\tilde A}_{J_1\cdots J_n}(\lambda_j, z_j,\bar z_j)$ on the celestial sphere demands:
\begin{align}\label{covariance}
\mathcal{\tilde A}_{J_1\cdots J_n}\left(\lambda_j,
{a z_j+b\over cz_j +d} ,{\bar a \bar z_j +\bar b\over\bar c\bar z_j+\bar d}\right)
  =  \prod_{j=1}^n \left[  
  (cz_j +d)^{\Delta_j  + J_j}
   (\bar c \bar z_j+\bar d)^{\Delta_j - J_j}  \right]
   \mathcal{\tilde A}_{J_1\cdots J_n}(\lambda_j,z_j,\bar z_j)\,,
\end{align}
as expected for a correlator of operators with weights $Delta_j$ and spins $J_j$.

\section{$n$-point MHV}
The cases of $3$- and $4$-point gluon amplitudes have been considered in \cite{Pasterski:2017ylz}. Here we will map $n\geq 5$-point MHV gluon amplitudes to the celestial sphere.

\subsection{Integrating out one $\omega_i$}

Starting from (\ref{mappingGen}), we can anchor the integration to one of our variables $\omega_i$  by 
making a change of variables for all $l\neq i$
\begin{align}
\omega_l \rightarrow  \frac{\omega_i}{s_i} \omega_l,
\end{align}
where $s_i$ is a constant factor that cancels the conformal scaling of $\omega_i$ in (\ref{confscale}), so that the ratio $\frac{\omega_i}{s_i}$ is conformally invariant.
One choice which is always possible in Minkowski signature is 
\begin{align}
\label{sidef}
s_i=\frac{|z_{i-1~i+1}|}{|z_{i-1~i}|\,|z_{i ~i+1}|}. 
\end{align}

Since gluon scattering amplitudes scale homogeneously under uniform rescalings, collecting all the factors in front,
we have 
\begin{align}
\label{alphaint}
\mathcal{\tilde A}_{J_1\cdots J_n}(\lambda_j, z_j,\bar z_j)
= \int_0^\infty \frac{d\omega_i}{\omega_i} \left(\frac{\omega_i}{s_i}\right)^{\sum_{j=1}^ni\lambda_j} s_i^{1+i\lambda_i}\left(\prod_{{a=1\atop a\neq i}}^n  \int_0^\infty d\omega_a  \, \omega_a^{i\lambda_a } \right)\,
\mathcal{A}_{\ell_1\cdots \ell_n} (s_i,\omega_l ,z_j,\bar z_j)\,,
\end{align}
where we used that the scaling power of dressed gluon amplitudes is $\mathcal{A}_n(\Lambda\omega_i)\to \Lambda^{-n}\mathcal{A}_n(\omega_i)$. We recognize that the integral over $\omega_i$ is the Mellin transform of $1$, which is given by
\begin{align}
 \int_0^\infty \frac{d\omega_i}{\omega_i} \left(\frac{\omega_i}{s_i}\right)^{iz} = 2\pi\delta(z) .
\end{align}
With this we simplify the transformation prescription (\ref{mappingGen}) to
\begin{align}
\label{finA}
\mathcal{\tilde A}_{J_1\cdots J_n}(\lambda_j, z_j,\bar z_j)
= 2\pi\delta \left(\sum_{j=1}^n\lambda_j \right)s_i^{1+i\lambda_i}\left(\prod_{{a=1\atop a\neq i}}^n  \int_0^\infty d\omega_a  \, \omega_a^{i\lambda_a } \right)\,
\mathcal{A}_{\ell_1\cdots \ell_n} (s_i,\omega_l,z_j\bar z_j)\,.
\end{align}

\subsection{Integrating out momentum conservation $\delta$-functions}




For simplicity, we choose the anchor variable above to be $\omega_1$ and use $\omega_{n-3}, \ldots, \omega_n$ to localize the momentum conservation $\delta$-functions in the amplitude.
These $\delta$-functions can then be equivalently rewritten as follows, compensating the transformation by a Jacobian:
\begin{align}\label{solng}
\delta^4(\epsilon_1 s_1q_1+\sum_{i=2}^n\epsilon_i\omega_i q_i)
&=\frac{4}{U}\prod_{j=n-3}^n s_j\delta\left(\omega_j-\omega_{j}^*\right)\mathbf{1}_{>0}(\omega_{j}^*),
\end{align}
where $\omega_{j}^*$ are solutions to the initial set of linear equations:
\begin{align}
\omega_j^\star =- s_j \left( \frac{U_{1,j}}{U}+ \sum_{i=2}^{n-4}\frac{\omega_i}{s_i}\frac{U_{i,j}}{U} \right).
\end{align}
The $U_{ij}$ and $U$ are minor determinants by Cramer's rule:
\begin{align}
U_{i,j}=\det(M^{\{n-3,...,j\to i,...,n\}}),~~~U=\det(M^{\{n-3,...,n\}}),
\end{align}
where $j\to i$ means that index $j$ is replaced by index $i$. $M^{\{a,b,c,d\}}$ denotes the $4\times 4$ matrix
\begin{align}
\label{matM}
M^{\{a,b,c,d\}}=(p_a\,p_b\,p_c\,p_d).
\end{align}
For the purpose of determinant calculation, the column vectors $p_i^\mu=\epsilon_i s_iq_i^\mu$ can be written in a manifestly conformally invariant form:
\begin{align}
\begin{split}
p_1^\mu (z, \bar z)  &=\epsilon_1 ( 1 , 0 , 0 , -1)~~,~~p_2^\mu (z, \bar z)  =\epsilon_2 ( 1 , 0 , 0 , 1)~~,~~p_3^\mu (z, \bar z)  =\epsilon_3 ( 2 , 2 , 0 , 0)\,,\\
p_i^\mu (z, \bar z)  &=\epsilon_i \frac{1}{|u_i|}( 1+|u_i|^2 , u_i+\bar u_i , -i (u_i-\bar u_i) , 1-|u_i|^2)~~~\text{for}~~~i=4,5,...,n\,,
\end{split}
\end{align}
in terms of conformal invariant cross-ratios
\begin{align}
\label{crossratios}
u_i = \frac{z_{31} z_{i2} }{ z_{32} z_{i1}} ~~\text{ and }~~ \bar u_i = \frac{\bar z_{31} \bar z_{i2} }{ \bar z_{32} \bar z_{i1}}~~~\text{for}~~~i=4,5,...,n\,,
\end{align}
but if, and only if, we also specify the explicit choice
\begin{align}
s_1=\frac{|z_{3,2}|}{|z_{3,1}|\,|z_{1,2}|}\,,~~s_2=\frac{|z_{3,1}|}{|z_{3,2}|\,|z_{2,1}|}\,,~~~\text{and}~~~s_i=\frac{|z_{1,2}|}{|z_{1,i}|\,|z_{i,2}|}~~~\text{for}~~~i=3,...,n.
\end{align}
The indicator functions $\prod_{i=n-3}^n\mathbf{1}_{>0}(\omega_{i}^*)$ appear due to the integration range in all $\omega$ being along the positive real line, such that the $\delta$-functions can only be localized in this region. 

Furthermore, in order for all the remaining integration variables $\omega_j$ with $j=2,...,n-4$ to be defined on the whole integration range, the indicator functions 
$\prod_{i=n-3}^n\mathbf{1}_{>0}(\omega_{i}^*)$ have to demand $\frac{U_{i,j}}{U}<0$ for all $i= 1,\ldots,n-4$ and $j=n-3,...,n$, so that we can write them as $\prod_{i,j}\mathbf{1}_{<0}(\frac{U_{i,j}}{U})$.

\subsection{Integrating the remaining $\omega_i$}

In this section we apply (\ref{finA}) to the usual $n$-point MHV Parke-Taylor amplitude \cite{Parke:1986gb} in spinor-helicity formalism for $n\geq 5$ rewritten via (\ref{spihel}):
\begin{align}
\mathcal{A}_{--+...+} (s_1,\omega_j,z_j,\bar z_j)=\frac{z_{12}^3s_1\omega_2 \delta^4(\epsilon_1s_1q_1+\sum_{i=2}^n\epsilon_i\omega_iq_i)}{(-2)^{n-4}z_{23}z_{34}...z_{n1}\omega_3\omega_4...\omega_n}.
\end{align}
Making use of the solutions (\ref{solng}) and performing four of the integrations in (\ref{finA}), we have:
\begin{align}
\mathcal{\tilde A}_{--+...+}(\lambda_i,z_i ,\bar z_i)&=2\pi\frac{\delta(\sum_{j=1}^n\lambda_j)z_{12}^3\,s_1^{i\lambda_1+2}}{(-2)^{n-4}U z_{23}z_{34}...z_{n1}}\prod_{a=2}^{n-4}  \int_0^\infty d\omega_a  \, \omega_a^{i\lambda_a }\frac{\omega_2 \prod_{b=n-3}^ns_{b}{\omega_{b}^*}^{i\lambda_{n-3}}}{\omega_3\omega_4...\omega_n^*}\prod_{i,j}\mathbf{1}_{<0}(\frac{U_{i,j}}{U}).
\end{align}
For convenience, we transform the remaining integration variables as:
\begin{align}
\omega_i=s_i\frac{U_{1,n}}{U_{i,n}}\frac{u_{i-1}}{1-\sum_{j=1}^{n-5}u_j}~~~,~~~i=2,3,...,n-4\,,
\end{align}
which leads to
\begin{align} \label{MHVcorrresult}
\mathcal{\tilde A}_{--+...+}(\lambda_i,z_i ,\bar z_i)\sim \frac{z_{12}^3s_1^{i\lambda_1+2}s_2^{i\lambda_2+2}s_{3}^{i\lambda_{3}}...s_{n}^{i\lambda_{n}}}{ z_{23}z_{34}...z_{n1}U_{1,n}}\delta(\sum_{j=1}^n\lambda_j)\,\hat\varphi(\{\alpha\},x)\, \prod_{i,j}\mathbf{1}_{<0}(\frac{U_{i,j}}{U}).
\end{align}
Note that the overall factor in (\ref{MHVcorrresult}) accounts for proper transformation weight of the resulting correlator under conformal transformations (\ref{covariance}). 

Here we recognize a hypergeometric function $\hat\varphi(\{\alpha\},x)$ of type $(n-4,n)$, as defined in section 3.8.1 of \cite{Aomoto:2011} and described in appendix \ref{hyperappendix}. In particular, here we have:
\begin{align}
\label{gluonPhi1}
\begin{split}
\hat\varphi(\{\alpha\},x)\equiv&\int_{\substack{u_1\geq 0,...,u_{n-5}\geq 0\\1-\sum_au_a\geq 0}} \prod_{j=1}^nP_j(u)^{\alpha_j}d\varphi~~~,~~~d\varphi=\frac{dP_{2}}{P_{2}}\wedge ... \wedge\frac{dP_{n-4}}{P_{n-4}}\,,\\
P_j(u)=&x_{0j}+x_{1j}u_1+...+x_{n-5\,j}u_{n-5}~~~,~~~1\leq j\leq n\,.
\end{split}
\end{align}
The parameters in (\ref{gluonPhi1}) corresponding to (\ref{MHVcorrresult}) read:\footnote{For $n=5$, the normally different cases $\alpha_2=2+i\lambda_2$ and $\alpha_{n-3}=i\lambda_{n-3}-1$ are reduced to a single $\alpha_2=1+i\lambda_2$. In this case there also are no integrations so that the result becomes a simple product of factors.}
\begin{align}
\alpha_1=&1\,,~\alpha_2=2+i\lambda_2\,,~\alpha_3=i\lambda_3\,,~...\,,~\alpha_{n-4}=i\lambda_{n-4}\,,~\alpha_{n-3}=i\lambda_{n-3}-1\,,~...\,,~\alpha_{n-1}=i\lambda_{n-1}-1,\notag\\
\label{gluparam}
\alpha_n=&1+i\lambda_1\,,~x_{0\,i}=\frac{U_{1,i}}{U_{1,n}}\,,~x_{j-1\,i}=\frac{U_{j,i}}{U_{j,n}}-\frac{U_{1,i}}{U_{1,n}}\,,~x_{0n}=-\frac{U}{U_{1,n}}\,,~x_{j-1\,n}=\frac{U}{U_{1,n}}\,,~x_{01}=1  \,,~x_{j-1\,j}=-\frac{U}{U_{j,n}}\,,
\end{align}
for $i=n-3,n-2,n-1$ and $j=2,3,...,n-4$, and all other $x_{ab}=0$.

These kinds of functions are also known as Aomoto-Gelfand hypergeometric functions on the Grassmannian $Gr(n-4,n)$. 

Making use of eq. (3.24) and (3.25) from \cite{Aomoto:2011}, we can write down a dual representation of the same function, which yields a hypergeometric function of type $(4,n)$:
\begin{align}
\label{gluonPhi2}
\begin{split}
\hat\varphi(\{\alpha\},x)\equiv&\frac{c_2}{c_1} \int_{\substack{u_1\geq 0,...,u_{3}\geq 0\\1-\sum_au_a\geq 0}} \prod_{j=1}^nP_j(u)^{\alpha_j}d\varphi~~~,~~~d\varphi=\frac{dP_{n-3}}{P_{n-3}}\wedge ... \wedge\frac{dP_{n-1}}{P_{n-1}}\,,\\
P_j(u)=&x_{0j}+x_{1j}u_1+x_{2j}u_2+x_{3j}u_3~~~,~~~1\leq j\leq n\,.
\end{split}
\end{align}
In this case, the parameters of (\ref{gluonPhi2}) corresponding to (\ref{MHVcorrresult}) read:
\begin{align}
\alpha_1=&1\,,~\alpha_2=-2-i\lambda_2\,,~\alpha_3=-i\lambda_3\,,...\,,~\alpha_{n-4}=-i\lambda_{n-4}\,,~\alpha_{n-3}=1-i\lambda_{n-3}\,,...\,,~\alpha_{n-1}=1-i\lambda_{n-1},\notag\\
\label{gluparam2}
\alpha_n=&-i\lambda_n\,,~x_{0j}=\frac{U_{j,n}}{U_{1,n}}\,,~x_{ij}=\frac{U_{j,n-4+i}}{U_{1,n-4+i}}-\frac{U_{j,n}}{U_{1,n}}\,,~x_{0n}=-\frac{U}{U_{1,n}}\,,~x_{in}=\frac{U}{U_{1,n}}\,,~x_{01}= 1   \,,\notag\\
x_{1\,n-3}=&\frac{-U}{U_{1,n-3}}\,,~x_{2\,n-2}=\frac{-U}{U_{1,n-2}}\,,~x_{3\,n-1}=\frac{-U}{U_{1,n-1}}\,,~\frac{c_2}{c_1}=\frac{\Gamma(2+i\lambda_1) \Gamma(2+i\lambda_2)\prod_{j=3}^{n-4}\Gamma(i\lambda_j)}{\Gamma(1-i\lambda_1)\prod_{i=1}^{3}\Gamma(1-i\lambda_{n-i})}.
\end{align}
for $i=1,2,3$ and $j=2,3,...,n-4$, and all other $x_{ab}=0$.

The hypergeometric functions $\hat\varphi(\{\alpha\},x)$ form a basis of solutions to a Pfaffian form equation which defines a Gauss-Manin connection as described in section 3.8 of \cite{Aomoto:2011}. This Pfaffian form equation can be interpreted as a generalized Knizhnik-Zamolodchikov equation satisfied by our correlators \cite{Abe:2015ucn,KNIZHNIK198483}.
Similar generalized hypergeometric functions appeared in \cite{Ferro:2014gca} in the context of $\mathcal{N}=4$ Yang-Mills scattering amplitudes
and the deformed Grassmannian.

\subsection{6-point MHV}
In the special case of six gluons there is only one integral in (\ref{gluonPhi1}), such that the function  reduces to the simpler case of Lauricella function $\hat\varphi_D$:
\begin{align}
\begin{split}
\hat\varphi_D(\{\alpha\},x)=&\left(\frac{-U}{U_{2,6}}\right)^{\text{i$\lambda $}_1+1}\left(\frac{-U}{U_{1,6}}\right)^{\text{i$\lambda $}_2+2} 
   \left(\frac{U_{2,3}}{U_{2,6}}\right)^{\text{i$\lambda $}_3-1} \left(\frac{U_{2,4}}{U_{2,6}}\right)^{\text{i$\lambda $}_4-1}
   \left(\frac{U_{2,5}}{U_{2,6}}\right)^{\text{i$\lambda $}_5-1}\times\\
	\label{lauricella}
	&\times\int_0^1dt\,t^{\alpha-1}(1-t)^{\gamma-\alpha-1}\prod_{i=1}^3(1-x_i t)^{-\beta_i},
\end{split}
\end{align}
with parameters and arguments given by
\begin{align}
\alpha=2+i\lambda_2,~\gamma=4+i\lambda_1+i\lambda_2,~\beta_i=1-i\lambda_{i+2},~x_i=1-\frac{U_{1,i+2} U_{2,6}}{U_{1,6} U_{2,i+2}}~\text{ for }~i=1,2,3.
\end{align}
Note that $x_{0j}$ arguments have been factored out of the integrand to achieve this form.

\section{$n$-point NMHV}
In this section we will map the $n$-point NMHV split helicity amplitude $\mathcal{A}_{---++\cdots +}$ to the celestial sphere via (\ref{finA}). The spinor-helicity expression for $\mathcal{A}_{---++\cdots+}$ can be found e.g. in \cite{Britto:2005dg}
\begin{align}
\mathcal{A}_{---++\cdots +} = \frac{1}{F_{3,1}} \sum_{j=4}^{n-1} \frac{\langle 1| P_{2,j} P_{j+1, 2} | 3 \rangle^3}{P_{2,j}^2 P_{j+1 , 2}^2} \frac{\langle j +1 ~ j \rangle}{[ 2 | P_{2,j} | j+1 \rangle \langle j | P_{j+1,2} | 2]} \equiv \sum_{j=4}^{n-1} \{ M_j \}
\end{align}
where $F_{i,j} \equiv \langle i ~ i+1 \rangle \langle i + 1 ~ i+2 \rangle \cdots \langle j-1 ~ j \rangle$ and $P_{x,y}  \equiv \sum_{k=x}^y | k \rangle [k|$ where $x<y$ cyclically. 

We will work with $\{M_4 \}$ for the purpose of our calculations. Using momentum conservation and writing $\{ M_4 \}$ in terms of spinor-helicity variables, we find
\begin{align}
\{ M_4 \} =&  \frac{1}{ \langle 34 \rangle \langle 45 \rangle \cdots \langle n-1 ~ n \rangle \langle n 1 \rangle}\frac{(\langle 1 2 \rangle [24] \langle 4 3 \rangle  + \langle 1 3 \rangle [3 4 ] \langle 4 3 \rangle ) ^3}{ (\langle 23 \rangle [23] + \langle 24 \rangle [24] + \langle 34 \rangle [34])  \langle 34 \rangle [34]}\times \notag\\
&\times \frac{\langle 54 \rangle}{([23] \langle 3 5 \rangle + [2 4 ] \langle 4 5 \rangle  )(\langle 43 \rangle [32])}.
\end{align}
Writing this in terms of celestial sphere variables via (\ref{spihel}), we find
\begin{align} \label{nptNMHVcelestial}
\{ M_4 \} = \frac{\frac{\omega_1 \omega _4 \left(\epsilon_2   z_{12} \bar{z}_{24} \omega _2 +\epsilon_3 z_{13}\bar{z}_{34} \omega _3 \right){}^3}{2^{n-4}  z_{56}  z_{67} \cdots z_{n-1,n} z_{n1} \bar{z}_{23} \bar{z}_{34}   \prod_{j=2, j\ne 4}^n \omega _j } }{  \left(\epsilon_3 z_{35} \bar{z}_{23}  \omega _3  +\epsilon_4  z_{45} \bar{z}_{24} \omega _4 \right) \left(\epsilon_2 \omega _2 \left(\epsilon_3   |z_{23}|^2 \omega _3 +\epsilon_4     |z_{24}|^2 \omega _4 \right)+\epsilon_3 \epsilon_4 |z_{34}|^2 \omega _3 \omega _4  \right)}.
\end{align}
The following map of the above formula to the celestial sphere will only be strictly valid for $n\geq 8$. We will comment on changes at 6- and 7-points in the next section. We use the map (\ref{finA}), anchor the calculation about $\omega_1$, make use of solutions (\ref{solng}) and perform a change of variables
\begin{align}
\omega_i=s_i\frac{u_{i-1}}{1-\sum_{j=1}^{n-5}u_j }, \quad i = 2, \ldots, n-4, 
\end{align}
to find the resulting term in the $n$-point NMHV correlator
\begin{align} \label{nptnmhvres}
\{ \tilde{M}_4 \} &\sim \delta \left(\sum_{j=1}^n\lambda_j \right) \frac{\prod_{i=1}^n s_i^{i \lambda_i}}{\bar{z}_{12} \bar{z}_{23} \bar{z}_{13} z_{45}z_{56}  \cdots z_{n-1, n}  z_{4,n}  } \frac{\bar{z}_{12} \bar{z}_{13}  z_{45} z_{4,n}  s_1^2 s_4^2    }{   \bar{z}_{34}  z_{n1}   U} \hat{\mathcal{F}} (\alpha, x) \prod_{i, j} {\bf 1}_{<0} (\frac{U_{i,j}}{U}),
\end{align}
with the function $\hat{\mathcal{F}} (\alpha, x)$ being a Gelfand $A$-hypergeometric function as defined in Appendix \ref{hyperappendix}. In this case it explicitly reads:
\begin{align}
\label{NMHVFres}
\begin{split}
\hat{\mathcal{F}} (\{\alpha\}, x)& =  \int_{{u_1 \geq 0,\ldots u_{n-5} \geq 0 \atop 1- u_1 - \cdots - u_{n-5} \geq 0} } \prod_{a=1}^{n-5} \frac{du_a}{u_a}   \, \prod_{j=1}^{n-5} u_j ^{i\lambda_{j+1}  } u_3 ^2  ( u_1 u_2 x_{10} + u_1u_3 x_{20} + u_2 u_3 x_{3 0}  )^{-1}  \\
&\times \prod_{i =1}^7 (x_{0i} + u_1 x_{1i} + \cdots +  u_{n-5} x_{n-5, i }  )^{\alpha_i},
\end{split}
\end{align}
where parameters are given by
\begin{align}
\alpha_1 = 3, ~ \alpha_2 = - 1,~ \alpha_3 = i \lambda_1 +1, ~ \alpha_4 = i \lambda_{n-3}-1 , ~ \alpha_5 = i \lambda_{n-2} -1,~\alpha_6 = i \lambda_{n-1} -1, ~ \alpha_7 = i \lambda_n -1,
\end{align}
and function arguments are given by
\begin{align}
x_{10} &= \epsilon_2 \epsilon_3   |z_{23}|^2  s_2 s_3, ~ x_{20} =  \epsilon_2 \epsilon_4     |z_{24}|^2 s_2 s_4 , ~ x_{30} = \epsilon_3 \epsilon_4 |z_{34}|^2 s_3 s_4, \notag \\
x_{11} &= \epsilon_2   z_{12} \bar{z}_{24} s_2, ~ x_{21} = \epsilon_3 z_{13}\bar{z}_{34} s_3 , ~ x_{22} = \epsilon_3 z_{35} \bar{z}_{23} s_3, ~ x_{32} =  \epsilon_4  z_{45} \bar{z}_{24} s_4, \notag  \\
x_{03} &= 1, ~ x_{j3} = -1, ~ j=1, \ldots , n-5, ~ x_{04} = \frac{U_{1,n-3}}{U} , \quad x_{j4} = \frac{U_{j, n-3} - U_{1,n-3}}{U}, ~ j = 1, \ldots, n-5, \notag  \\
x_{05} &= \frac{U_{1,n-2}}{U} , \quad x_{j5} = \frac{U_{j, n-2} - U_{1,n-2}}{U}, ~ j = 1, \ldots, n-5, \\
x_{06} &= \frac{U_{1,n-1}}{U} , \quad x_{j6} = \frac{U_{j, n-1} - U_{1,n-1}}{U}, ~ j = 1, \ldots, n-5, \notag \\
x_{07} &= \frac{U_{1,n}}{U} , \quad x_{j7} = \frac{U_{j, n} - U_{1,n}}{U}, ~ j = 1, \ldots, n-5. \notag  
\end{align} 
Note that the first fraction in (\ref{nptnmhvres}) accounts for the correct transformaton weight of the correlator under conformal tranformation (\ref{covariance}).

\subsection*{ 6- and 7-point NMHV}

In the cases of 6- and 7-point the results in the previous section change somewhat, due to the presence of $\omega_3$ and $\omega_4$ in the denominator of (\ref{nptNMHVcelestial}). These variables are fixed by momentum conservation $\delta$-functions in the lower point cases, such that the parameters and function arguments of the resulting Gelfand $A$-hypergeometric functions change. 

For the 6-point case, we find that the resulting correlator part $\{ \tilde{M}_4 \}$ is proportional to a Gelfand $A$-hypergeometric function as defined in Appendix \ref{hyperappendix}:
\begin{align}
\label{nmhv6pt}
\begin{split}
\hat{\mathcal{F}} (\{\alpha\}, x)& =  \int_{{u_1 \geq 0 \atop 1- u_1 \geq 0} } \frac{du_1}{u_1}   \, u_1 ^{i\lambda_2  } ( x_{00} +  u_1 x_{10} +u_1^2 x_{2 0}  )^{-1} (1 - u_1  )^{i\lambda_1+1} \prod_{i =2}^7 (x_{0i} + u_1 x_{1i}  )^{\alpha_i} 
\end{split}
\end{align}
where parameters are given by
\begin{align}
 \alpha_2 = i \lambda_3 - 1, ~ \alpha_3 = i \lambda_4 + 1, ~ \alpha_4 = i \lambda_5 - 1, ~ \alpha_5 = i \lambda_6 - 1, ~  \alpha_6 = 3, ~ \alpha_7 = -1, 
\end{align}
and function arguments $x_{ij}$ depend on $\epsilon_i,z_i,\bar z_i$ and $U_{ij}$.
Performing a partial fraction decomposition on the quadratic denominator in (\ref{nmhv6pt}), we can reduce the result to a sum of two Lauricella functions.

In the 7-point case, we find that the resulting correlator part $\{ \tilde{M}_4 \}$ is proportional to a Gelfand $A$-hypergeometric function as defined in Appendix \ref{hyperappendix}:
\begin{align}
\begin{split}
\hat{\mathcal{F}} (\{\alpha\}, x)& =  \int_{{u_1 \geq 0, u_2 \geq 0 \atop 1- u_1 - u_2 \geq 0} } \frac{du_1}{u_1} \frac{d u_2}{u_2}  \, u_1 ^{i\lambda_2  } u_2 ^{i \lambda_3 } ( u_1 x_{10} + u_2 x_{20} + u_1 u_2 x_{3 0} + u_1^2 x_{4 0} + u_2^2 x_{50} )^{-1}  \\
&\times \prod_{i =1}^7 (x_{0i} + u_1 x_{1i} + u_2 x_{2i }  )^{\alpha_i},
\end{split}
\end{align}
where parameters are given by
\begin{align}
\alpha_1 &= i \lambda_1 + 1,~\alpha_2 = i \lambda_4 + 1, ~ \alpha_3 = i \lambda_5 -1,~ \alpha_4 = i \lambda_6 -1,~\alpha_5 = i \lambda_7 - 1, ~ \alpha_6 = 3,~\alpha_7 = -1,
\end{align}
and function arguments $x_{ij}$ again depend on $\epsilon_i,z_i,\bar z_i$ and $U_{ij}$.

\section{$n$-point N${}^k$MHV }

In this section we discuss the schematic structure of N${}^k$MHV amplitudes with higher $k$ under the Mellin transform (\ref{finA}).

\subsection*{N${}^2$MHV amplitude}

In the 8-point N${}^2$MHV split helicity case, $\mathcal{A}_{----++++}$, we consider one of the six terms of  the amplitude found in e.g. \cite{Britto:2005dg} on page 6 as an example:
\begin{align}\label{8ptn2mhvspihel}
\frac{1}{F_{4,1} \bar{F}_{2,3}} \frac{\langle 1 | P_{2,6} P_{7,2} P_{3,5}  P_{6,3}| 4 \rangle^3}{P_{2,6}^2 P_{7,2}^2P_{3,5}^2 P_{6,3}^2 } \frac{\langle 76 \rangle [23]\langle 65 \rangle }{[2 | P_{2,6} | 7 \rangle \langle 6 | P_{7,2} | 2] [3 | P_{3,5} | 6 \rangle \langle 5 | P_{6,3} | 3] },
\end{align}
where $\bar{F}_{i,j}$ is the complex conjugate of ${F}_{i,j}$. Performing the same sequence of steps as in the previous sections, we find a resulting Gelfand $A$-hypergeometric function of the form
\begin{align}
&\hat{\mathcal{F}} (\{\alpha\}, x) =  \int_{{u_1 \geq 0, u_2 \geq 0, u_3 \geq 0 \atop 1- u_1 - u_2 - u_3 \geq 0} } \frac{du_1}{u_1} \frac{d u_2}{u_2}  \frac{du _3 }{u_3} \, u_1 ^{\alpha_1  } u_2 ^{\alpha_2 } u_3^{\alpha_3}  \mathcal{P}_{\{4\}}^3 \prod_{i =4}^{13} (x_{0i} + u_1 x_{1i} + u_2 x_{2i } + u_3 x_{3i}  )^{\alpha_i} \\
&\times \prod_{j=14}^{17} (x_{0j} + u_1 x_{1j} + u_2 x_{2j} + u_3 x_{3j} + u_1 u_2 x_{4j} + u_1 u_3 x_{5j} + u_2 u_3 x_{6j} + u_1^2 x_{7j} + u_2 ^2 x_{8j} + u_3^2 x_{9j})^{\alpha_j},\notag
\end{align}
for some parameters $\alpha_i$, where $\mathcal{P}_{\{4\}}$ is a degree four polynomial in $u_i$, and function arguments $x_{ij}$ again depend on $\epsilon_i,z_i,\bar z_i$ and $U_{ij}$.

\subsection*{N${}^k$MHV amplitude}

More generally a split helicity N${}^{k}$MHV amplitude $\mathcal{A}_{-\cdots-+\cdots+}$ involves a sum over the terms described in eq. (3.1), (3.2) of \cite{Britto:2005dg}. Terms corresponding in complexity to $\{ \tilde{M}_4 \}$ discussed in the previous section are always present, with constant Laurent polynomial powers at any $k$. However, for higher $k$, the most complicated contributing summands result in hypergeometric integrals schematically given by
\begin{align}
\label{NkMHVres}
 \hat{\mathcal{F}}(\{\alpha\},x)=& \int_{{u_1, \ldots, u_{n-4} \geq 0 \atop 1 - u_2 - \cdots - u_{n-4} \geq 0}}  \prod_{{l=2}}^{n-4} \frac{d u_l}{u_l} u_l^{\alpha_l}\, \left(1- \sum_{j=2}^{n-4 } u_j \right)^{\alpha_1 }  \mathcal{ P}_{\{2k\}}^3 \left(\prod_i(\mathcal{ P}_{\{1\}}^{i})^{\alpha_i}\right)\left(\prod_j(\mathcal{ P}_{\{2\}}^j)^{\alpha_j}\right)
\end{align}
where $\alpha_i$ are parameters and $\mathcal{ P}_{\{d\}}$ is a degree $d$ polynomial in $u_a$. Here we explicitly see an increase in power of the Laurent polynomials with increasing $k$ in N${}^{k}$MHV. The examples above feature the Gelfand $A$-hypergeometric function $\hat{\mathcal{F}}$. The increase in Laurent polynomial degree is traced back to the presence of Mandelstam invariants $P_{i,j}^2$ for degree two polynomials, as well as the factors $\langle a|P_{i,j}P_{k,l}...P_{r,t}|b \rangle$ for higher degree polynomials. The length of chains of the $P_{i,j}$ depends on $n$ and $k$, such that multivariate Laurent polynomials of any positive degree are present at sufficiently high $n,k$.

Similar generalized hypergeometric functions, or, equivalently, generalized Euler integrals are found in the case of string scattering amplitudes \cite{Oprisa:2005wu, Mafra:2011nw}. It will be interesting to explore this connection further.

\acknowledgments

We have benefitted greatly from
 discussions with Shu-Heng Shao, Marcus Spradlin and James Stankowicz.
We thank Congkao Wen and Dhritiman Nandan for collaboration on related work \cite{Nandan:2018}.
This work was supported in part by the US Department of Energy under contract
DE-SC0010010 Task A,
Simons Investigator Award \#376208 and 
the IBM Einstein Fellowship (AV).

\appendix
\section{Generalized hypergeometric functions}
\label{hyperappendix}
The Aomoto-Gelfand hypergeometric functions of type $(n+1,m+1)$ relevant in this work can be defined as in section 3.5.1 of \cite{Aomoto:2011}:
\begin{align}
\label{phinm}
\hat\varphi(\{\alpha\},x)\equiv&\int_{\substack{u_1\geq 0,...,u_{n}\geq 0\\1-\sum_au_a\geq 0}} \prod_{j=0}^mP_j(u)^{\alpha_j}d\varphi\,,\\
d\varphi=&\frac{dP_{j_1}}{P_{j_1}}\wedge ... \wedge\frac{dP_{j_n}}{P_{j_n}}\,~~~,~~~0\leq j_1<...<j_n\leq m\,,\\
\label{Pfactor}
P_j(u)=&x_{0j}+x_{1j}u_1+...+x_{nj}u_{n}~~~,~~~1\leq j\leq m\,,
\end{align}
where here the parameters $\alpha_i$ collectively describe all the powers for the factors in the integrand. When all $\alpha_i$ are zero, the function reduces to the Aomoto polylogarithm.

The arguments $x_{ij}$ of the hypergeometric function of type $(m+1,n+1)$ in (\ref{Pfactor}) can be arranged in a matrix:
\begin{align}
\label{Xbar}
\bar X=\left(\begin{array}{ccccccccccc}
      x_{00} & \dots & x_{0m}  \\
      x_{10} & \dots  & x_{1m}  \\
   \vdots & \ddots & \vdots  \\
     x_{n0} & \dots  & x_{nm} 
\end{array}\right).
\end{align}

Each column in this matrix defines a hyperplane in $\mathbb{C}^n$ that appears in the hypergeometric integral as $(x_{0j}+\sum_{i=1}^nx_{ij}u_i)^{\alpha_i}$. Furthermore, $(n+1)\times(n+1)$ minor determinants of the matrix can be regarded as Pl\"ucker coordinates on the Grassmannian $Gr(n+1,m+1)$ over the space of arguments $x_{ij}$. 

Sometimes it is convenient to transform the argument arrangement (\ref{Xbar}) to the following gauge fixed form
\begin{align}
\left(\begin{array}{ccccccccccc}
    1 & 0  & \dots &  0  & 1 & 1 & \dots & 1  \\
    0 &  1 & \dots &  0  & -1 & -x_{11} & \dots  & -x_{1\,m-n-1}  \\
\vdots&    & \ddots &    & -1 & \vdots & \vdots & \vdots  \\
    0 & 0 & \dots  & 1  & -1 & -x_{n1} & \dots  & -x_{n\,m-n-1} 
\end{array}\right).
\end{align}
In this case the hypergeometric function can then be written in the following two equivalent ways, eq. (3.24) of \cite{Aomoto:2011}:
\begin{align}
\label{hypernm}
F((\alpha_i),(\beta_j),\gamma;x)=&c_1\int_{\substack{u_1\geq 0,...,u_{n}\geq 0\\1-\sum_au_a\geq 0}} d^nu\prod_{i=1}^nu_i^{\alpha_i-1}\cdot (1-\sum_{l=1}^nu_l)^{\gamma-\sum_i\alpha_i-1}\prod_{j=1}^{m-n-1}(1-\sum_{i=1}^nx_{ij}u_i)^{-\beta_j},\notag\\
c_1=&\Gamma(\gamma)/\Gamma(\gamma-\sum_{i=1}^n\alpha_i)\cdot\prod_{i=1}^n\Gamma(\alpha_i),
\end{align}
and the dual representation in eq. (3.25) of \cite{Aomoto:2011}:
\begin{align}
\label{hypernmDual}
F((\alpha_i),(\beta_j),\gamma;x)=&c_2\int_{\substack{u_1\geq 0,...,u_{m-n-1}\geq 0\\1-\sum_au_a\geq 0}} d^{m-n-1}u\prod_{i=1}^{m-n-1}u_i^{\beta_i-1}\cdot (1-\sum_{l=1}^{m-n-1}u_l)^{\gamma-\sum_i\beta_i-1}\prod_{j=1}^{n}(1-\sum_{i=1}^{m-n-1}x_{ji}u_i)^{-\alpha_j},\notag\\
c_2=&\Gamma(\gamma)/\Gamma(\gamma-\sum_{i=1}^{m-n-1}\beta_i)\cdot\prod_{i=1}^{m-n-1}\Gamma(\beta_i),
\end{align}
where the parameters are assumed to satisfy the conditions
\begin{align}
\begin{split}
&\alpha_i\notin\mathbb{Z},~1\leq i\leq n,~\beta_j\notin \mathbb{Z},~1\leq j\leq m-n-1,\\
&\gamma-\sum_{i=1}^n\alpha_i\notin \mathbb{Z},~\gamma-\sum_{j=1}^{m-n-1}\beta_j\notin\mathbb{Z}.
\end{split}
\end{align}

The hypergeometric functions (\ref{phinm}) comprise a basis of solutions to the defining set of differential equations
\begin{align}
(1)~~~&\sum_{i=0}^n x_{ij}\frac{\partial \hat\varphi}{\partial x_{ij}}=\alpha_j \hat\varphi \, ,&&~ 0\leq j\leq m,\notag\\
(2)~~~& \sum_{j=0}^m x_{ij}\frac{\partial \hat\varphi}{\partial x_{ij}}=-(1+\alpha_i) \hat\varphi \, ,&&~ 0\leq i\leq n,\\
(3)~~~& \frac{\partial^2 \hat\varphi}{\partial x_{ij}\partial x_{pq}}=\frac{\partial^2 \hat\varphi}{\partial x_{iq}\partial x_{pj}},&&~0\leq i,p\leq n,~~0\leq j,q\leq m.\notag
\end{align}

In cases where factors of the integrand are non-linear in the integration variables, the functions can be generalized further to Gelfand $A$-hypergeometric functions \cite{Gel'fand1986,Gel'fand1990255} defined as:
\begin{align}
\hat{\mathcal{F}}(\{\alpha\},x)=\int_{\substack{u_1\geq 0,...,u_{k}\geq 0\\1-\sum_au_a\geq 0}}\prod_{i}\mathcal{P}_i(u_1,...,u_k)^{\alpha_i}u_1^{\alpha_1}...u_k^{\alpha_k} du_1...du_k,
\end{align}
where $\alpha_i$ are complex parameters and $\mathcal{P}_i$ now are Laurent polynomials in $u_1,...,u_k$. 

%
%
%

%
%

\begin{thebibliography}{99}

\bibitem{flat} 
  J.~de Boer and S.~N.~Solodukhin,
  ``A Holographic reduction of Minkowski space-time,''
  Nucl.\ Phys.\ B {\bf 665}, 545 (2003)
  doi:10.1016/S0550-3213(03)00494-2
  [hep-th/0303006].
   T.~Banks,
  ``The Super BMS Algebra, Scattering and Holography,''
  arXiv:1403.3420 [hep-th].
   A. ~Ashtekar, ``Asymptotic Quantization: Based On 1984 Naples
Lectures,`` Naples, Italy: Bibliopolis,(1987).
   C.~Cheung, A.~de la Fuente and R.~Sundrum,
  ``4D scattering amplitudes and asymptotic symmetries from 2D CFT,''
  JHEP {\bf 1701}, 112 (2017)
  doi:10.1007/JHEP01(2017)112
  [arXiv:1609.00732 [hep-th]].
  D.~Kapec, P.~Mitra, A.~M.~Raclariu and A.~Strominger,
  ``2D Stress Tensor for 4D Gravity,''
  Phys.\ Rev.\ Lett.\  {\bf 119}, no. 12, 121601 (2017)
  doi:10.1103/PhysRevLett.119.121601
  [arXiv:1609.00282 [hep-th]].
  D.~Kapec, V.~Lysov, S.~Pasterski and A.~Strominger,
  ``Semiclassical Virasoro symmetry of the quantum gravity $
\mathcal{S}$-matrix,''
  JHEP {\bf 1408}, 058 (2014)
  doi:10.1007/JHEP08(2014)058
  [arXiv:1406.3312 [hep-th]].
  F.~Cachazo and A.~Strominger,
  ``Evidence for a New Soft Graviton Theorem,''
  arXiv:1404.4091 [hep-th].
    A.~Strominger,
  ``Lectures on the Infrared Structure of Gravity and Gauge Theory,''
  arXiv:1703.05448 [hep-th].
  	

\bibitem{Pasterski:2016qvg} 
  S.~Pasterski, S.~H.~Shao and A.~Strominger,
  ``Flat Space Amplitudes and Conformal Symmetry of the Celestial Sphere,''
  arXiv:1701.00049 [hep-th].

\bibitem{Cardona:2017keg} 
  C.~Cardona and Y.~t.~Huang,
  ``S-matrix singularities and CFT correlation functions,''
  JHEP {\bf 1708}, 133 (2017)
  doi:10.1007/JHEP08(2017)133
  [arXiv:1702.03283 [hep-th]].

\bibitem{Nandan:2018} 
D.~Nandan, A.~Volovich, C.~Wen, and M.~Zlotnikov, work in progress.

\bibitem{Pasterski:2017kqt} 
  S.~Pasterski and S.~H.~Shao,
  ``A Conformal Basis for Flat Space Amplitudes,''
  arXiv:1705.01027 [hep-th].

\bibitem{Pasterski:2017ylz} 
  S.~Pasterski, S.~H.~Shao and A.~Strominger,
  ``Gluon Amplitudes as 2d Conformal Correlators,''
  arXiv:1706.03917 [hep-th].
  
  
  
  
\bibitem{Lam:2017ofc} 
  H.~T.~Lam and S.~H.~Shao,
  ``Conformal Basis, Optical Theorem, and the Bulk Point Singularity,''
  arXiv:1711.06138 [hep-th].
  
\bibitem{Banerjee:2017jeg} 
  N.~Banerjee, S.~Banerjee, S.~A.~Bhatkar and S.~Jain,
  ``Conformal Structure of Massless Scalar Amplitudes Beyond Tree level,''
  arXiv:1711.06690 [hep-th].


\bibitem{other}
F.~Cachazo, S.~He and E.~Y.~Yuan,
  ``Scattering of Massless Particles in Arbitrary Dimensions,''
  Phys.\ Rev.\ Lett.\  {\bf 113}, no. 17, 171601 (2014)
  doi:10.1103/PhysRevLett.113.171601
  [arXiv:1307.2199 [hep-th]].
N.~Arkani-Hamed, F.~Cachazo, C.~Cheung and J.~Kaplan,
  ``A Duality For The S Matrix,''
  JHEP {\bf 1003}, 020 (2010)
  doi:10.1007/JHEP03(2010)020
  [arXiv:0907.5418 [hep-th]].
  R.~Roiban, M.~Spradlin and A.~Volovich,
  ``On the tree level S matrix of Yang-Mills theory,''
  Phys.\ Rev.\ D {\bf 70}, 026009 (2004)
  doi:10.1103/PhysRevD.70.026009
  [hep-th/0403190].
N.~Arkani-Hamed, F.~Cachazo, C.~Cheung and J.~Kaplan,
  ``The S-Matrix in Twistor Space,''
  JHEP {\bf 1003}, 110 (2010)
  doi:10.1007/JHEP03(2010)110
  [arXiv:0903.2110 [hep-th]].
N.~Arkani-Hamed, Y.~Bai and T.~Lam,
  ``Positive Geometries and Canonical Forms,''
  JHEP {\bf 1711}, 039 (2017)
  doi:10.1007/JHEP11(2017)039
  [arXiv:1703.04541 [hep-th]].
	
	
\bibitem{Srednicki:2007qs} 
  M.~Srednicki,
  ``Quantum field theory,''
	
\bibitem{Parke:1986gb} 
  S.~J.~Parke and T.~R.~Taylor,
  ``An Amplitude for $n$ Gluon Scattering,''
  Phys.\ Rev.\ Lett.\  {\bf 56}, 2459 (1986).
  doi:10.1103/PhysRevLett.56.2459

\bibitem{Aomoto:2011} 
  K.~Aomoto, M.~Kita,
  ``Theory of Hypergeometric Functions,'' (Springer-Verlag, Tokyo, 2011), DOI:\href{https://link.springer.com/book/10.1007\%2F978-4-431-53938-4}{10.1007/978-4-431-53938-4}, ISBN 9784431539384.

\bibitem{KNIZHNIK198483}
  V.~G.~Knizhnik and A.~B.~Zamolodchikov, 
  ``Current algebra and Wess-Zumino model in two dimensions",
  Nucl. Phys. B., 247 1 (1984), pp. 83-103,
  doi:10.1016/0550-3213(84)90374-2.


  
  
\bibitem{Abe:2015ucn} 
  Y.~Abe,
  ``A note on generalized hypergeometric functions, KZ solutions, and gluon amplitudes,''
  Nucl.\ Phys.\ B {\bf 907}, 107 (2016)
  doi:10.1016/j.nuclphysb.2016.03.032
  [arXiv:1512.06476 [hep-th]].

  
\bibitem{Ferro:2014gca} 
  L.~Ferro, T.~Lukowski and M.~Staudacher,
  ``$\mathcal N=4$ scattering amplitudes and the deformed Grassmannian,''
  Nucl.\ Phys.\ B {\bf 889}, 192 (2014)
  doi:10.1016/j.nuclphysb.2014.10.012
  [arXiv:1407.6736 [hep-th]].

\bibitem{Britto:2005dg} 
  R.~Britto, B.~Feng, R.~Roiban, M.~Spradlin and A.~Volovich,
  ``All split helicity tree-level gluon amplitudes,''
  Phys.\ Rev.\ D {\bf 71}, 105017 (2005)
  doi:10.1103/PhysRevD.71.105017
  [hep-th/0503198].

\bibitem{Oprisa:2005wu} 
  D.~Oprisa and S.~Stieberger,
  ``Six gluon open superstring disk amplitude, multiple hypergeometric series and Euler-Zagier sums,''
  hep-th/0509042.
   
   
   \bibitem{Mafra:2011nw} 
  C.~R.~Mafra, O.~Schlotterer and S.~Stieberger,
  ``Complete N-Point Superstring Disk Amplitude II. Amplitude and Hypergeometric Function Structure,''
  Nucl.\ Phys.\ B {\bf 873}, 461 (2013)
  doi:10.1016/j.nuclphysb.2013.04.022
  [arXiv:1106.2646 [hep-th]].




\bibitem{Gel'fand1986}
I.~M.~Gel'fand, 
``General theory of hypergeometric functions", 
Dokl. Akad. Nauk SSSR 288 (1986), no. 1, 14-18. 



	


\bibitem{Gel'fand1990255}
  I.~M.~Gel'fand and M.~M.~Kapranov and A.~V.~Zelevinsky, 
  ``Generalized Euler integrals and A-hypergeometric functions",
  Adv. Math., 84 2 (1990), pp. 255-271,
  doi:10.1016/0001-8708(90)90048-R.


	


\end{thebibliography}


%

\end{document}